\documentclass[ps,prd,preprintnumbers,superscriptaddress,nofootinbib,floatfix,twocolumn,notitlepage]{revtex4-2}
\pdfoutput=1 

\usepackage{dcolumn}

\usepackage{bm,amssymb,slashed,graphicx,multirow,soul,mathtools,xspace,array,tikz,amsmath,siunitx}
\usepackage[compat=1.1.0]{tikz-feynman}   
\usepackage{float}                   
\allowdisplaybreaks
\usepackage{ bbold }
\usepackage{caption,subcaption}
\usepackage{braket}
\usepackage{hyperref}

\definecolor{nicered}{rgb}{0.7,0.1,0.1}
\definecolor{nicegreen}{rgb}{0.1,0.5,0.1}
\definecolor{violet}{rgb}{0.7,0.3,0.3}
\hypersetup{colorlinks,citecolor= nicegreen,linkcolor= nicered}

\newcommand{\lp}{\left(}
\newcommand{\rp}{\right)}

\newcommand{\beq}{\begin{equation} }
\newcommand{\eeq}{\end{equation}} 
\newcommand{\bi}{\begin{itemize} }
\newcommand{\ei}{\end{itemize} }

\definecolor{Red}{rgb}{1.,0.,0.}
\definecolor{Grn}{rgb}{0.,0.75,0.}
\definecolor{Blu}{rgb}{0.,0.,1.}
\definecolor{Pink}{rgb}{1,0.08,0.58}

\usepackage[T1]{fontenc} 

\usepackage{amsmath,amssymb,epsfig,ulem,color,slashed}
\allowdisplaybreaks  

\newcommand{\Xm}{X_{\rm max}}
\newcommand{\gcm}{~{\rm g}/{\rm cm}^2} 

\begin{document} 


\title{\boldmath Ultra High Energy Cosmic Rays versus Models of High Energy Hadronic Interactions }

\author{Bla\v{z} Bortolato}
\email{blaz.bortolato@ijs.si}
\author{Jernej F. Kamenik}
\email{jernej.kamenik@cern.ch}
\affiliation{
 Jo\v{z}ef Stefan Institute, Jamova 39, 1000 Ljubljana, Slovenia \\
 Faculty of Mathematics and Physics, University of Ljubljana, Jadranska 19, 1000 Ljubljana, Slovenia
}

\author{Michele Tammaro}
\email{michele.tammaro@fi.infn.it}
\affiliation{INFN Sezione di Firenze, Via G. Sansone 1, I-50019 Sesto Fiorentino, Italy
}

\date{\today}

\begin{abstract}

We evaluate the consistency of hadronic interaction models in the CORSIKA simulation package with publicly available fluorescence telescope data from the Pierre Auger Observatory. By comparing the first few central moments of the extended air shower depth maximum distributions, as extracted from measured events, to those predicted by the best-fit inferred compositions, we derive a statistical measure of the consistency of a given hadronic model with data. To mitigate possible systematic biases, we include all primaries up to iron, compensate for the differences between the measured and simulated energy spectra of cosmic rays and account for other known systematic effects. Additionally, we study the effects of including higher central moments in the fit and project our results to larger statistics.
 
\end{abstract}

\maketitle

\section{Introduction}
\label{sec:intro}

The study of Ultra-High Energy Cosmic Rays (UHECRs), particles with energies exceeding $10^{18}$ eV incident on Earth, is crucial for understanding the most energetic processes in the Universe. Key questions in UHECR research include their origin, propagation mechanisms, and composition. The origin of UHECRs is still debated, with potential sources ranging from active galactic nuclei to gamma-ray bursts~\cite{Bertone:2002ks, Horandel:2006xf}. Understanding their propagation involves considering interactions with cosmic microwave background photons and (inter)galactic magnetic fields, which crucially affect their trajectories and energy spectra. Additionally, the composition of UHECRs, whether primarily protons, or a mixture of heavier nuclei, has significant implications for both their sources, acceleration and propagation~\cite{Andringa:2011zz}.

Detection of UHECRs currently relies on large ground based observatories such as the Pierre Auger Observatory~\cite{PierreAuger:2015eyc, Stasielak:2021hjm} and the Telescope Array~\cite{KAWAI2008221}, comprised of water Cherenkov detectors and fluorescence telescopes. They measure various observables related to extended air showers (EAS) produced by UHECRs, from which the energy, arrival direction, and the composition of primary cosmic rays can be inferred. 

Interpreting the data from these observatories requires detailed simulations of EAS initiated by UHECRs in the Earth's atmosphere. These simulations depend crucially on models of hadronic interactions, which describe the behavior of particles produced in high-energy collisions. The accuracy of these models is essential for reliable interpretation of experimental data.

Recent developments in hadronic interaction models~\cite{Ostapchenko:2024myl,PierreAuger:2023htc,Blazek:2021jzf, Ostapchenko:2024vjo}, have improved our understanding of particle interactions at extreme energies. These models are continually calibrated and refined using data from both cosmic ray experiments and particle accelerators. 

However, some discrepancies persist, particularly in the observed ground-level muon multiplicity, which appears inconsistent with predictions~\cite{Cazon:2020zhx}. Conversely, measurements from fluorescence telescopes show better agreement with simulations, highlighting a possible venue for further refinement of hadronic interaction models.

Our recent work has focused on developing novel observables, such as the central moments of the shower depth maximum ($\Xm$) distributions recorded by the fluorescence detectors, to infer the mass composition of UHECRs~\cite{Bortolato:2022ocs, Bortolato:2023wsk, Bortolato:2024npt}. These observables demonstrated differentiating sensitivity to hadronic models, providing a potential pathway to resolve current discrepancies and enhance our understanding of UHECR composition and interactions.

In this work, we carefully reexamine the consistency of different hadronic models with $\Xm$ measurements. Previously~\cite{Bortolato:2022ocs}, we have shown that for the limited statistics of the Pierre Auger Open Data, the first three $\Xm$ central moments suffice to infer the full UHECR mass composition. 
Here we extend this study by projecting to larger statistics, relevant for the analysis of the full Pierre Auger dataset. In particular, we focus on the sensitivity of higher $\Xm$ moments to hadronic interaction models. By marginalizing over possible UHECR mass compositions we construct a robust statistical measure to differentiate between and assess the validity of different models.

The remainder of the article is structured as follows. In Section~\ref{sec:methods} we review the inference procedure for UHECR mass composition, with the respective likelihood based on $\Xm$ moments, and discuss the interplay of the CR energy spectrum, mass composition and hadronic interaction model effects. In Section~\ref{sec:results} we present and discuss our results on hadronic model discrimination, while our main conclusions are summarized in Section~\ref{sec:Conclusions}.

\section{Methodology}
\label{sec:methods}

Following our previous work~\cite{Bortolato:2022ocs}, we infer the full mass composition\footnote{We define the full composition as $w=(w_p,\dots,w_{Fe})$, that is with a total of 26 primaries, from proton ($p$) to iron ($Fe$). For a recent discussion on the choice of composition components see ref.~\cite{Bortolato:2024npt}.} of UHECR through the first $n$ central moments of the $X_{\max}$ distribution. The statistical method based on bootstrapping allows to include in a natural way both systematic uncertainties from detector effects, and statistical errors from the finite size of data and simulated events available. Moreover, it provides a natural framework to make projections to larger statistics, in case the number of events is increased by a multiplicative factor $f$~\cite{Bortolato:2023wsk}, as explained below. 

\subsection{PA Open Data}

We use publicly available data from the 2021 Pierre Auger Open Data (PAOD) release~\cite{the_pierre_auger_collaboration_2021_4487613}, consisting of $\sim$10\% of the entire dataset collected by the Pierre Auger Observatory. In order to study $\Xm$, we are restricted to the use of hybrid showers, that is events recorded by both the Surface Detectors (SD) and the Fluorescent Detectors (FD); furthermore, we restrict our analysis to the energy bin $E\in[0.65, 1]$ EeV, containing 934 events. In this way we can select a statistical significant sample while limiting possible systematic errors from the energy dependence of the observables when comparing to simulations, as discussed  in more details below. 
Systematic uncertainties on the measured $\Xm$ distributions have as main sources the detector calibration and the reconstruction of the data, as detailed in Ref.~\cite{PierreAuger:2014sui}. These amount to an error $\sim10\gcm$ for all energies. Larger uncertainties from the detector resolution are instead of statistical nature; fluctuations in the number of photo-electrons detected and in the shower arrival directions lead to a $\sim25\gcm$ systematic uncertainty on the observed $\Xm$ at low energies, while it decreases to $\sim15\gcm$ at higher energies. These values are expected to decrease in the future, as the larger dataset collected will improve the detector resolution spread.

\subsection{Comparing with EAS Simulations}

In parallel, we simulate the $\Xm$ distributions of EAS generated by different primaries with CORSIKA v7.7550~\cite{Heck:1998vt}. For simplicity, we assume a uniform energy distribution in the bin $E\in[0.65,1]$ EeV; then, for each of the 26 primaries between proton ($Z=1$) and iron ($Z=26$), and for each of the hadronic models (EPOS~\cite{Pierog:2013ria}, Sibyll~\cite{Riehn:2024prp}, QGSJetII-04~\cite{Ostapchenko:2010vb} and QGSJet01~\cite{Kalmykov:1993qe})\footnote{Recently, new implementations of these models are being developed: EPOS4~\cite{Werner:2024fwk}, Sibyll$^\bigstar$~\cite{Riehn:2024prp} and QGSJetIII~\cite{Ostapchenko:2024myl}. However, their predictions of EAS observables are expected to be consistent with those of previous models included in this work.}, we simulate 2000 showers, focusing on their longitudinal electromagnetic energy profiles. 
Note that by including all primaries (up to iron) we avoid potential systematics associated with the choice of primary components~\cite{Bortolato:2024npt}. 
Finally, we include systematic uncertainties in the simulated $\Xm$ distributions. The smearing from the detector resolution can be added following the procedure in Ref.~\cite{PierreAuger:2014sui}. Note that this effect is statistics limited and that the fit of the smearing parameters is based on the current PA dataset. We do not attempt to estimate the expected improvement on these parameters from the present and future data collection, thus projections to statistics beyond the present PA data can be considered conservative.  

Another potentially important source of systematic uncertainty comes from the energy spectra of primaries, since it is well known that $\Xm$ distributions exhibit distinct primary energy dependence. Thus simulations based on a uniform energy distribution cannot be directly compared with PA data following steeply falling spectra. Furthermore, current EAS energy measurements only determine the energy spectrum of all primaries combined. The indications of an energy dependent composition of UHECRs however point towards distinct energy spectra of light and heavier primaries, as one would expect if different mechanisms are responsible for the creation and/or acceleration of light and heavy primaries.

\begin{figure*}[t!]
    \centering
    \includegraphics[scale = 0.7]{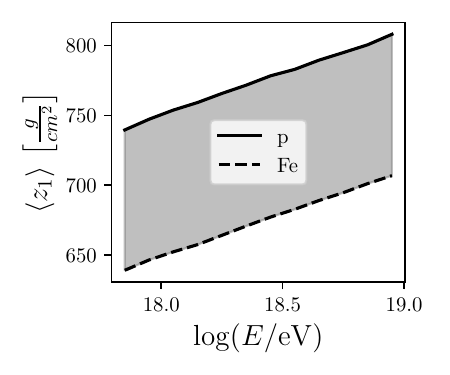}
    \includegraphics[scale = 0.7]{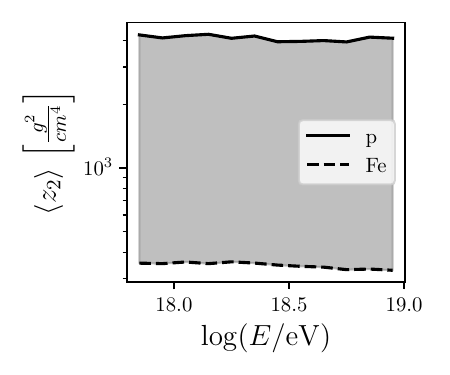}
    \includegraphics[scale = 0.7]{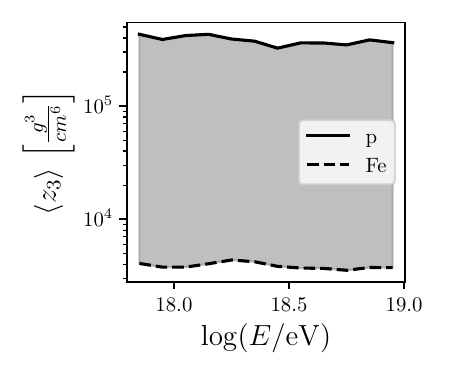}
    \caption{The averaged value of the moments $z_1$ (top left), $z_2$ (top right) and $z_3$ (bottom), for proton (solid line) and Fe (dashed line),  as a function of $\log_{10}(E/{\rm{eV}})$. Simulated using the EPOS hadronic model.}
    \label{fig:z1logE}
\end{figure*}

To illustrate this point, we plot in Fig.~\ref{fig:z1logE} the energy dependence of the first three central moments computed using the EPOS hadronic model for protons (solid lines) and iron (dashed lines).
We observe the well known logarithmic dependence of $z_1  \equiv \langle \Xm \rangle$ while higher moments do not show any significant energy dependence on these scales. We also note that even for $z_1$, the energy dependence is small compared to the total range of values between light and heavy primaries. In absence of any energy dependence, the composition inferred from data in a finite energy interval would simply be the average composition. The energy dependence of $z_1$ thus introduces a systematic bias.  In order to determine whether it significantly affects the inferred composition and its likelihood, we assume in the following that the spectra of all primaries are identical within the chosen narrow energy interval of 0.65~EeV to 1~EeV. Even in this case, the uniform energy spectrum of the simulated data for each primary differs from the measured data. We can however account for this bias by reweighing the measured events so that their spectrum matches the one simulated (flat).
We can then compare inferred compositions with and without reweighing to estimate the significance of this bias.
The weights can be assigned to each measured event as the inverse of the energy spectrum density of the event itself, $u_i = 1/P(E_i)$. The energy spectrum density can be estimated in various ways, such as binning, template (i.e. power law) fitting, or using techniques like kernel density estimation. In this work, we estimate $P(E)$ by binning PAOD events. We split the relevant energy interval into 30 bins of equal length and compute the binned energy spectrum density by counting all events in a bin ($N_b$) and dividing by the total number of all events ($N$), $P(E)_b= N_b/N$.\footnote{With the chosen binning, the number of events per energy bin, $N_b$, ranges between 20 and 60 for the interval $E\in [0.65,1]$ EeV.} For each event, we then determine the corresponding energy bin and assign it a weight equal to the inverse of the energy spectrum density of that bin, $1/P(E_i) \simeq 1/P(E)_b$.

\subsection{Probing Hadronic Models}

\begin{figure*}
    \centering
    \includegraphics[scale = 0.9]{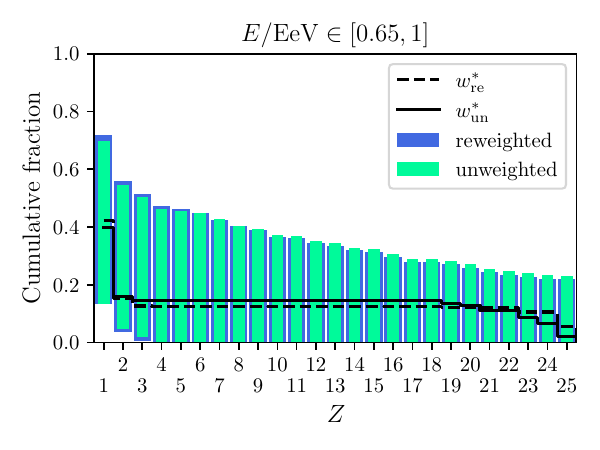}\includegraphics[scale = 0.9]{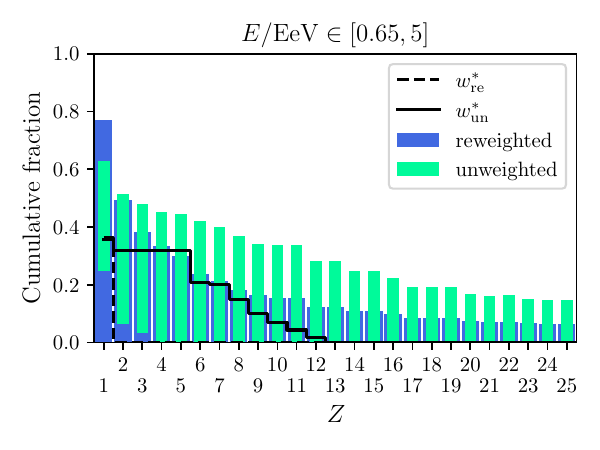}
    \caption{Best fit of the cumulative composition and the respective confidence intervals, in two energy bins, $[0.65,1]$ EeV ({\bf Left}) and  $[0.65,5]$ EeV ({\bf Right}), obtained with the EPOS hadronic model. The solid black line and green bars indicate the results obtained without introducing any weights for the energy distribution of measured events, while the dashed black line and blue bars show the results derived by including the reweighting procedure described in the text.}
    \label{fig:eReweighting:Comparison}
\end{figure*}

To compare how well a specific hadronic model describes the measured data, considering $n$ central moments, we first compute the best composition for the case at hand. We start from the log-likelihood function
\beq\label{eq:likelihood}
\ln{\cal L}(w) = \int \ln\Big[ p\lp z| w \rp \Big]~ P\lp z \rp ~{\rm d}z\,,
\eeq
where $p\lp z| w \rp$ is the probability distribution function (PDF) of moments $z = (z_1,..., z_n)$, computed from simulated data for a given composition $w$;
similarly, $P(z)$ is the PDF of moments $z$, computed from measured data (PAOD).

These distributions are not well-determined due to systematic and statistical uncertainties of the measured and simulated events, which in turn affect the $\Xm$ moments, the energy weights and ultimately the mass composition. We account for these uncertainties via the bootstrap method~\cite{EfroTibs93}.
Namely, we use random uniform sampling, with repetitions allowed, to compute moments of the distributions, and repeat the procedure for $B = 10^4$ times. For a fixed energy bin, which includes $N$ showers, the procedure is as follows:
\begin{enumerate}
    \item Randomly choose $N$ events, with uniform probability and allowing for repetitions. We denote the selected events as $\{ (\tilde{E}_i, \tilde{X}_{\max,i}, \delta \tilde{X}_{\max,i}) \}_\ell$, with $i =1,\dots, N$. Here $\ell$ is the current bootstrap step, while $\delta\tilde{X}_{\max,i}$ is the total systematic uncertainty of $\tilde{X}_{\max,i}$.
    
    \item Estimate the energy density, $P(\tilde E_i)$, and compute the weight for each event, $\{ u_i \} = 1/P(\tilde E_i)$.
    
    \item Compute the moments:
    
    \begin{align}
        z^\ell_1 &= \frac{\sum^N_{i=1} \left( \tilde{X}_{\max,i} + \delta \tilde{X}_{\max,i} \, \epsilon_i \right) u_i}{\sum^N_{i=1} \, u^l_i} \,,
        \label{eq:Moments:Mean} \\
        z^\ell_n &= \frac{\sum^N_{i=1} \left( \tilde{X}_{\max,i} + \delta \tilde{X}_{\max,i} \, \epsilon_i - z^l_1 \right)^n \, u_i}{\sum^N_{i=1} u_i} \,,
        \label{eq:Moments:Higher}
    \end{align}
    
    where $\epsilon_i$ are samples from a normal distribution with mean $0$ and standard deviation $1$.

    \item Repeat the previous steps for $\ell=1,\dots,B$.
\end{enumerate}
The procedure to compute central moments from data remains the same as described in Ref.~\cite{Bortolato:2022ocs}. 

For simulated events, this procedure needs to be carried out for each primary separately, with $N=N_{\rm sim}=2000$ the total number of events in the simulated dataset; the moments of each primary then receive a different weight depending on the composition $w$~\cite{Bortolato:2022ocs}.
When analyzing a real dataset instead, $N$ corresponds to the actual number of observed events in the chosen energy bin; in our case, we have $N=N_{\rm PAOD}=934$ for hybrid events with $E_i \in[0.65, 1]$ EeV in PAOD. The bootstrap then yields a distribution of observed moments, $P(z)$, to which we fit the distribution for simulated events to infer $w$. The variance of $P(z)$ strongly depends on the available statistics, that is $N_{\rm PAOD}$; we can make projections to larger statistics by sampling $f$-times the number of events in the actual PAOD dataset, where $f$ is a statistical multiplicative factor, such that $N= f N_{\rm PAOD}$. Concretely, this factor represents an effective reduction in the variance of $P(z)$ that would follow a larger statistical sample, assuming the central value remains unchanged.

Due to the bootstrap method, both $p(z|w)$ and $P(z)$ follow a multivariate normal distribution; that is $p(z | w) = \mathcal{N}_n(\mu_s, \Sigma_s)$ and  $P(z) = \mathcal{N}_n(\mu_m, \Sigma_m)$, where $\mu_s$ and $\mu_m$ ($\Sigma_s$ and $\Sigma_m$) indicate the vector of means (covariance matrix) obtained from simulations and measurements, respectively. 
The most probable composition, $w^*$, can then be computed by maximizing the log-likelihood in eq.~\eqref{eq:likelihood}; given the properties of $p(z|w)$ and $P(z)$ discussed above, the integral can be computed analytically~\cite{Bortolato:2022ocs}. To estimate the confidence level of the resulting $w^*$, we use Nested Sampling techniques to efficiently sample the log-likelihood around its maximum~\cite{doi:10.1063/1.1835238, Buchner_2014, https://doi.org/10.48550/arxiv.1707.04476, https://doi.org/10.48550/arxiv.2101.09604}. Note that in general these results depend on both the number of moments included, $n$, and the statistical multiplier assumed, $f$.

Finally, we compute the confidence level of rejecting a given hadronic model, as function of $n$ and $f$, by comparing the distribution of moments for the best fit composition, $p(z|w^*)$, to the measured one, $P(z)$. As the difference between random variables of two multivariate normal distributions also follows a multivariate normal distribution, $z^s - z^m \sim \mathcal{N}(\mu_s - \mu_m, \Sigma_s + \Sigma_m)$, we can compute the rejection power by evaluating the expression
\beq\label{eq:chisquared:moments}
\chi^2_n = (\mu_s - \mu_m)^T (\Sigma_s + \Sigma_m)^{-1} (\mu_s - \mu_m),
\eeq
which follows a $\chi^2$ distribution with $n$ degrees of freedom.

\section{Results}
\label{sec:results}

\begin{figure*}[t!]
    \centering
    \includegraphics[scale = 0.85]{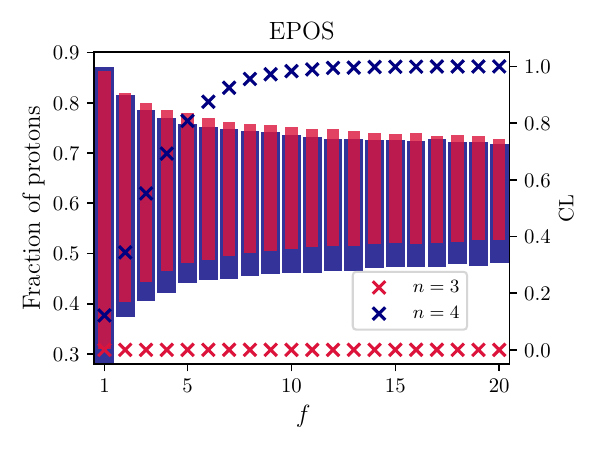}
    \includegraphics[scale = 0.85]{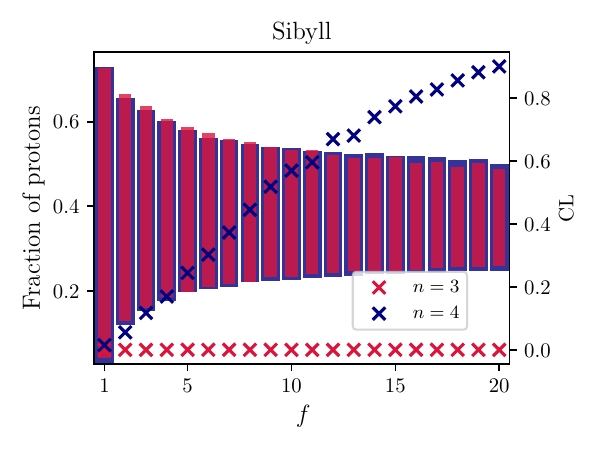}
    \caption{Inferred fraction of protons (left axis) and model CL of rejection (right axis) as function of the statistics multiplier $f$, using EPOS (left plot) and Sibyll (right plot). Red and blue bars show the proton fraction results with $n=3$ and $n=4$ respectively, while crosses refer to the CL, with the same color code. }
    \label{fig:ConfidenceLevel:EPOS&SIB}
\end{figure*}

We first test the sensitivity of our approach to the size of the chosen energy interval by comparing the inferred most probable composition (and its likelihood) when accounting for, or not, the energy spectrum of measured events (i.e. using weights $u_i=1/P(E_i)$ or $u_i=1$, respectively). We find no significant difference between the inferred composition with original or energy reweighted events for the interval $E_i\in [0.65,1]$~EeV, see left panel of Fig.~\ref{fig:eReweighting:Comparison}. This justifies the choice of the energy interval size, indicating that the described systematic bias does not significantly impact the results. On the other hand, selecting a larger energy bin would amplify this effect, as shown in the right panel of Fig.~\ref{fig:eReweighting:Comparison}, where we consider all events in the energy bin $[0.65,5]$ EeV. To have a quantitative estimate of the associated systematic bias, we can compute the (energy reweighted) log-likelihood ($\log \mathcal L_{\rm re}$) ratio of the most probable compositions as inferred using reweighted ($w^*_{\rm re}$) and unweighted ($w^*_{\rm un}$) fits, assuming some statistical multiplier $f$. Namely, we compute
\beq
\lambda(f) = \frac{\log{\cal L}_{\rm re}(w^*_{\rm un}, f)}{\log{\cal L}_{\rm re}(w^*_{\rm re}, f)}\,.
\eeq
In the small $E_i\in[0.65,1]$ EeV bin, we obtain $\lambda(f=1)\simeq0.99$ with the PAOD statistics, and only decreases to $\lambda(f=5)\simeq0.97$. Conversely, for the choice of a larger energy bin, $E_i\in[0.65,5]$ EeV, the log-likelihood ratio is already $\lambda(f=1)\simeq0.29$ with current statistics, and goes down to $\lambda(f=5)\simeq0.1$ assuming a dataset five times larger. We thus conclude, that while the effect is certainly important for large energy intervals, it is completely negligible for narrow enough bins, both in current PAOD as well as in anticipated larger statistics event samples.

The main result of this work is shown in Fig.~\ref{fig:ConfidenceLevel:EPOS&SIB}, choosing the EPOS and Sibyll hadronic models as case studies.\footnote{We present the same results for the two QGSJet models in Appendix~\ref{app:QGS}.} The fraction of protons inferred with $n = 3$ (red) and $n=4$ (blue) moments is shown as a function of $f$. As expected, the determination of $w_p$ becomes more and more precise with increasing $f$, that is with increasing statistics, while the introduction of an additional moment in the inference procedure has a weak effect (see also Ref.~\cite{Bortolato:2022ocs}). At the same time, we show on the right y-axis the rejection confidence levels, with crosses that follow the same color code as for the bars, computed with eq.~\eqref{eq:chisquared:moments}. While the $n=3$ case is fully consistent with the PAOD across the entire shown range of $f$, the inclusion of the fourth moment yields a rejection CL that is monotonically increasing. At $f=1$, that is the actual PAOD, we have $CL\simeq 20\%$, which however grows rapidly, reaching $CL\gtrsim 90\%$ already at $f\lesssim 10$. In other words, already with the existing full PA dataset and by including additional information into the inference procedure, the fit obtained with the EPOS model could be statistically inconsistent with the data. A similar behaviour can be observed with the Sibyll hadronic model, see the right plot in Fig.~\ref{fig:ConfidenceLevel:EPOS&SIB}, although the rejection level grows more slowly with $f$.

\begin{figure}[t]
    \centering
    \includegraphics[scale = 0.85]{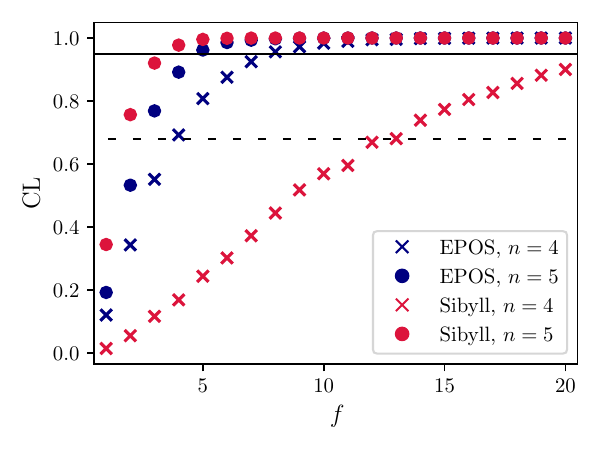}
    \caption{Rejection CL of EPOS (blue) and Sibyll (red) as function of $f$, using $n=4$ (cross) and $n=5$ (dots) moments. The horizontal solid (dashed) line indicates the $68\%$ ($95\%$) confidence level.}
    \label{fig:CLwith4and5:EPOS&SIB}
\end{figure}

To detail further the effect of the number of moments and statistics on the rejection of hadronic models, we show in Fig.~\ref{fig:CLwith4and5:EPOS&SIB} the results for CL by adding an additional moment to the likelihood, that is $n=5$, for both EPOS and Sibyll cases. Clearly, the rate of growth of model rejection confidence level increases significantly with the inclusion of an extra observational constraint, reaching $CL\gtrsim95\%$ already at $f=5$ for both models.

\begin{figure*}[t!]
    \centering
    \includegraphics[scale = 0.8]{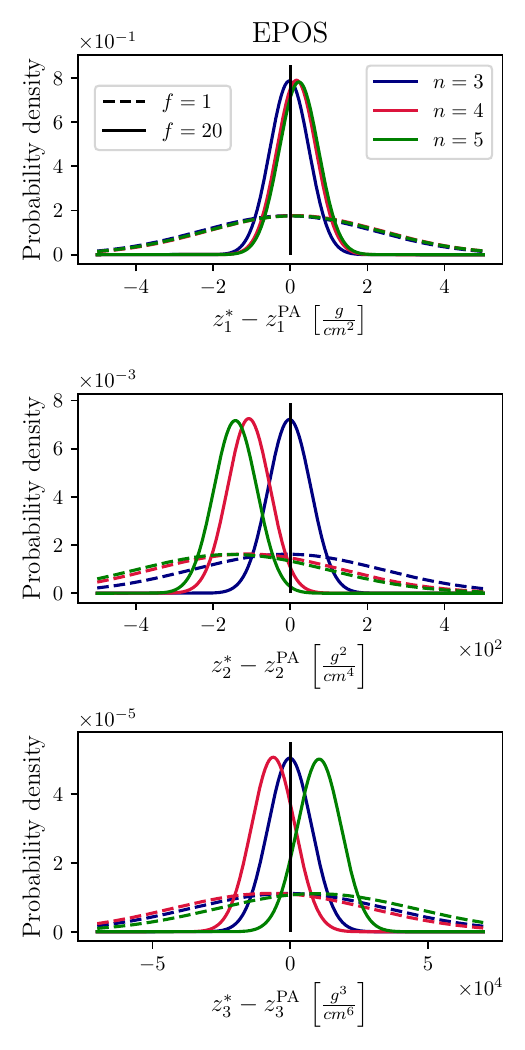}
    \includegraphics[scale = 0.8]{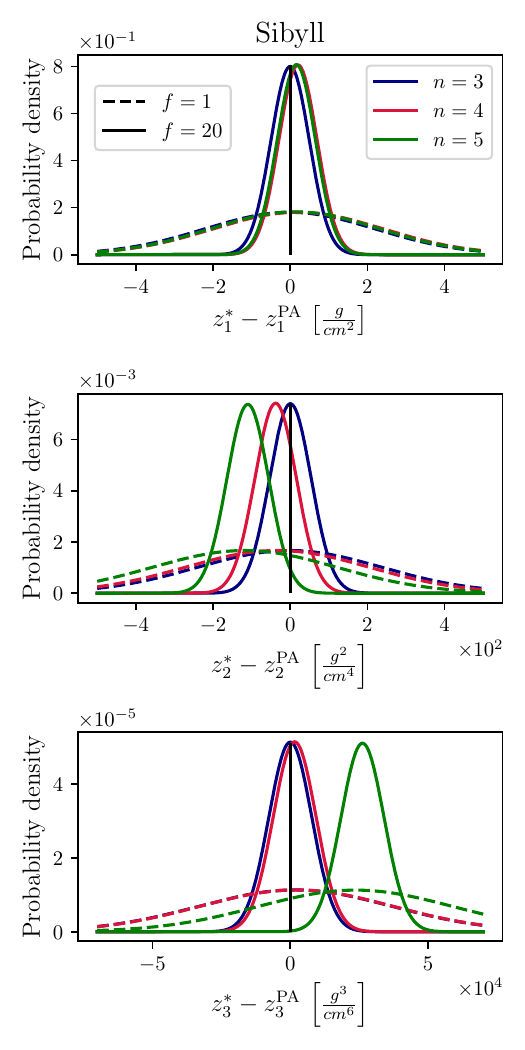}
    \caption{Deviation of the first three moments of the best fit composition, $z_i^{*}$, from the PAOD moments, $z_i^{\rm PA}$, using the EPOS (left column) and Sibyll (right column) models. Blue, red and green indicate the use of $n=3,4,5$ moments respectively, while solid and dashed lines refer to $f=1$ and $f=20$.}
    \label{fig:MomentsDifference:EPOS}
\end{figure*}

Finally, we can explicitly compute the deviations of the single best fit moments from the respective PAOD measured moments and thus identify the effects of including higher moments on the individual pulls in the likelihood fit. In Fig.~\ref{fig:MomentsDifference:EPOS} we show the first three moments of the best fit composition, $z_i^*$ with $i=1,2,3$, obtained with the EPOS model, with respect to the PAOD moments, $z_i^{\rm PA}$. We show the results for three different choices of $n$, $n=3,4,5$, and two choices of $f$, $f=1,20$. The top plot, referring to the first moment $z_1$, indicates that the mean of the inferred $\Xm$ distribution is consistent with the data (the vertical black line) for all the cases considered. That is, in general the first moment is well reconstructed, even with several higher moments included in the fit and at much larger statistics than currently available. In contrast, the best fits for $z_2$ and $z_3$ start deviating from data once we include more moments, with a clear enhancement of the discrepancy when projecting to larger statistics. In the future, such analysis could thus point to particular tensions within fits to hadronic models failing to simultaneously account for all included $\Xm$ distribution moments.

\section{Conclusions}
\label{sec:Conclusions}

This work examines the consistency of hadronic interaction models implemented in the CORSIKA simulation package (EPOS, Sibyll, QGSJetII-4 and QGSJet01), with the publicly available fluorescence telescope data measured at the Pierre Auger Observatory. For each model, the first $n$ moments of the $\Xm$ distribution extracted from measured events are compared to the moments given by the best (most probable) inferred compositions, including all primaries up to iron and after accounting for all other known possible systematic effects. A statistical measure of consistency based on this comparison then allows us to compute confidence levels of rejecting a particular hadronic model. The results are furthermore projected to (future) larger datasets by assuming to have $f$ - times more measured (and simulated) events. 

As a byproduct of our analysis, the systematics associated with the unknown primary CR spectra and finite energy bin intervals are considered. We show that the effects of the overall CR energy spectrum can be accounted for by properly reweighing the measured events included in the fit to match the spectrum of simulated events. We also find that these effects are in practice negligible for narrow enough energy bins, for which the inferred composition can thus be interpreted as averaged over the relevant energy interval.

For EPOS and Sibyll models, we find that the best fit compositions are consistent with currently available public data. As long as we restrict the inference procedure to the first $n=3$ moments the conclusion persists even when projecting to much larger datasets (up to $f=20$). 
Once additional moments are introduced however, the rejection CL grows with projected increased statistics, highlighting a possible incompatibility of hadronic model predictions with measured data. Since our projections are based on fixed central values of measured moments, this does not mean that these models will in fact be incompatible once more data is analyzed. Instead it demonstrates the sensitivity of the existing full PA dataset (and its future upgrades) to discriminate between hadronic interaction models that currently best describe the available data.\footnote{For the QGSJetII and QGSJet01 models, we show in Appendix~\ref{app:QGS} that already the first and the second moment, respectively, lead to poor fits to measured data.} 

The method presented here can hopefully lay the foundations for more in-depth studies to test and improve hadronic interaction models employed in simulations of UHECR initiated atmospheric shower development. In turn this could shed light on existing discrepancies observed in ground detector data as well as improve our confidence in the primary composition spectra already inferred from measurements with fluorescence telescopes.

\section*{Acknowledgments}
B.B. and J.F.K. acknowledge the financial support from the Slovenian Research Agency (grant No. J1-3013 and research core funding No. P1-0035).

\bibliography{references}

\begin{appendix}
\onecolumngrid

\section{Results for QGSJet models}
\label{app:QGS}

\begin{figure}[h!]
    \centering
    \includegraphics[scale = 0.85]{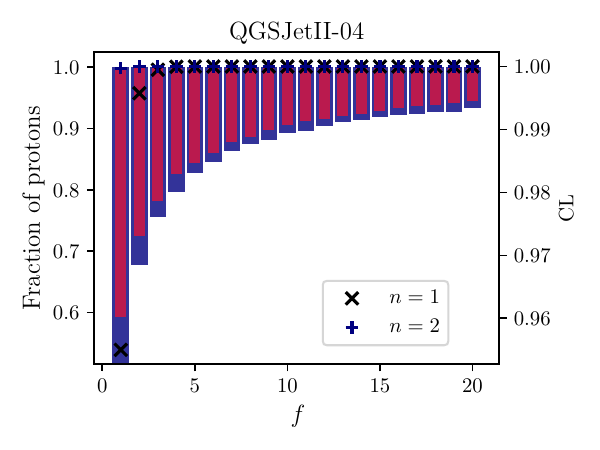}
    \includegraphics[scale = 0.85]{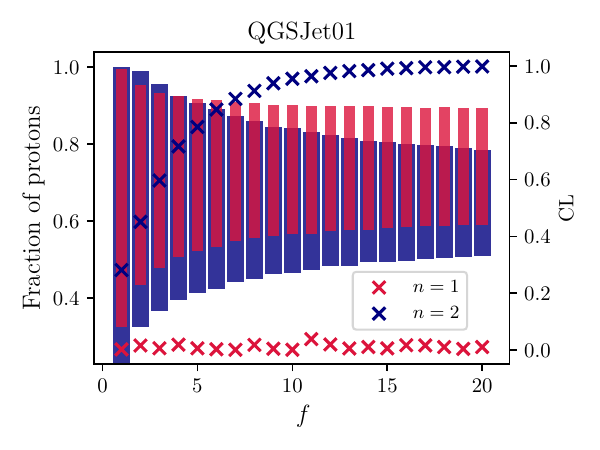}
    \caption{Inferred fraction of protons (left axis) and model CL of rejection (right axis) as function of the statistics multiplier $f$, using QGSJetII-04 (left plot) and QGSjet01 (right plot). Red and blue bars show the proton fraction results with $n=1$ and $n=2$ respectively, while crosses refer to the CL, with the same color code except for QGSJetII-04.}
    \label{fig:ConfidenceLevel:QGS}
\end{figure}

\begin{figure}[h!]
    \centering
    \includegraphics[scale = 0.85]{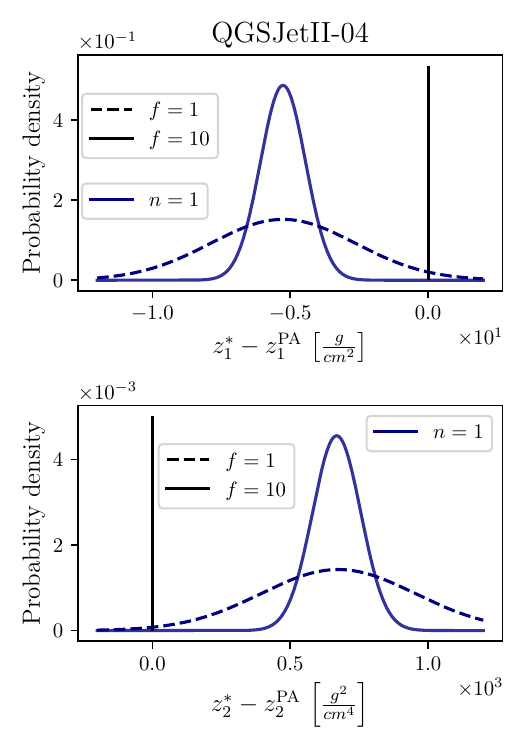}\hspace{1cm}
    \includegraphics[scale = 0.85]{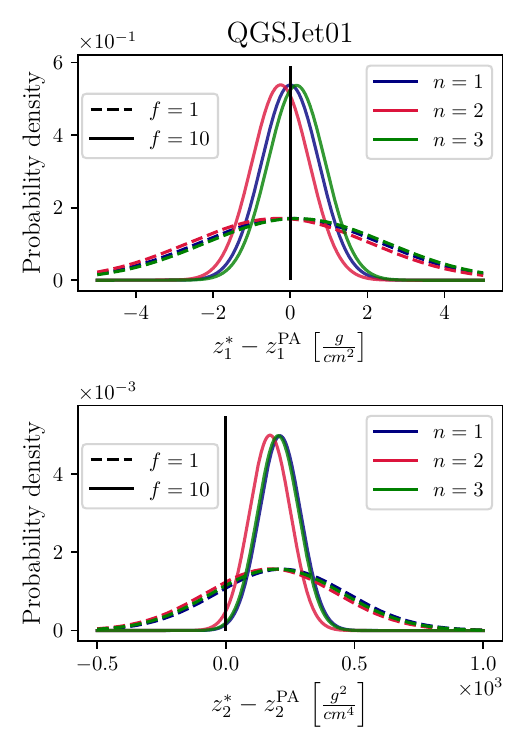}
    \caption{Deviation of the first two moments of the best fit composition, $z_i^{*}$, from the PAOD moments, $z_i^{\rm PA}$, using the QGSJetII-04 (left column) and QGSJet01 (right column) models. Blue, red and green indicate the use of $n = 1,2,3$ moments respectively, while solid and dashed lines refer to $f=1$ and $f=10$. 
    For QGSJetII-04 the deviation of the first and second moment are the same for $n=1,2,3$.
    }
    \label{fig:QGSJetMomentsDeviation}
\end{figure}

\begin{figure}
    \centering
    \includegraphics[scale = 0.85]{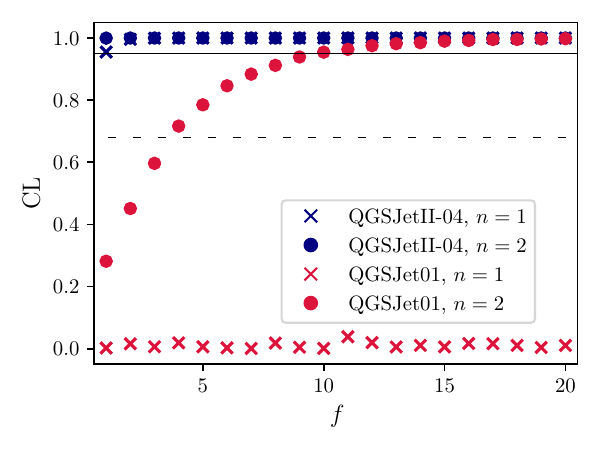}
    \caption{Rejection CL of QGSJetII-04 (blue) and QGSJet01 (red) as function of $f$, using $n=1$ (cross) and $n=2$ (dots) moments. The horizontal solid (dashed) line indicates the $68\%$ ($95\%$) confidence level.}
    \label{fig:enter-label}
\end{figure}

\end{appendix}

\end{document}